\documentclass[useAMS,usenatbib]{mn2e}
\usepackage{epsfig}
\usepackage{deluxetable} 
\usepackage{lineno}
\usepackage{hyperref}

\title[The most powerful flaring activity from the NLSy1 PMN\,J0948$+$0022]{The most powerful flaring activity from the NLSy1
  PMN\,J0948$+$0022} \author[F. D'Ammando, M. Orienti, J. Finke, et
al.]{F. D'Ammando$^{1,2}$\thanks{\emph{Fermi} LAT Collabration Member}\thanks{E-mail: dammando@ira.inaf.it},
  M. Orienti$^{2}$\footnotemark[1], J. Finke$^{3}$\footnotemark[1],
  C. M. Raiteri$^{4}$, T. Hovatta$^{5}$, J. Larsson$^{6}$\footnotemark[1],
  \newauthor W. Max-Moerbeck$^{7}$\footnotemark[1],
  J. Perkins$^{8}$\footnotemark[1], A. C. S. Readhead$^{5}$\footnotemark[1],
  J. L. Richards$^{9}$\footnotemark[1], and \newauthor 
M.~Beilicke$^{10}$, W.~Benbow$^{11}$, K.~Berger$^{12}$, R.~Bird$^{13}$,
V.~Bugaev$^{10}$, J.~V~Cardenzana$^{14}$, \newauthor M.~Cerruti$^{11}$,
X.~Chen$^{15,16}$, L.~Ciupik$^{17}$, H.~J.~Dickinson$^{14}$,
J.~D.~Eisch$^{14}$, M.~Errando$^{18}$, \newauthor
A.~Falcone$^{19}$, J.~P.~Finley$^{9}$, H.~Fleischhack$^{16}$, P.~Fortin$^{11}$,
L.~Fortson$^{20}$, A.~Furniss$^{21}$, \newauthor
L.~Gerard$^{16}$, G.~H.~Gillanders$^{22}$, S.~T.~Griffiths$^{23}$,
J.~Grube$^{17}$, G.~Gyuk$^{17}$, N.~H{\aa}kansson$^{15}$, \newauthor  
J.~Holder$^{12}$, T.~B.~Humensky$^{24}$, P.~Kar$^{25}$, M.~Kertzman$^{26}$,
Y.~Khassen$^{13}$, D.~Kieda$^{25}$,\newauthor
F.~Krennrich$^{14}$, S.~Kumar$^{12}$, M.~J.~Lang$^{22}$,
G.~Maier$^{16}$, A.~McCann$^{27}$,  K.~Meagher$^{28}$, \newauthor
P.~Moriarty$^{22}$, R.~Mukherjee$^{18}$, D.~Nieto$^{24}$, A.~O'Faol\'{a}in de Bhr\'{o}ithe$^{16}$,
R.~A.~Ong$^{29}$, \newauthor
A.~N.~Otte$^{28}$, M.~Pohl$^{15,16}$, A.~Popkow$^{29}$, H.~Prokoph$^{16}$,
E.~Pueschel$^{13}$, J.~Quinn$^{13}$, \newauthor
K.~Ragan$^{30}$, P.~T.~Reynolds$^{31}$, G.~T.~Richards$^{28}$,
E.~Roache$^{11}$, J.~Rousselle$^{29}$, \newauthor M.~Santander$^{18}$,
G.~H.~Sembroski$^{9}$, A.~W.~Smith$^{25}$, D.~Staszak$^{30}$,
I.~Telezhinsky$^{15,16}$, \newauthor
J.~V.~Tucci$^{9}$, J.~Tyler$^{30}$, A.~Varlotta$^{9}$, 
V.~V.~Vassiliev$^{29}$, S.~P.~Wakely$^{32}$, A.~Weinstein$^{14}$, \newauthor
R.~Welsing$^{16}$, D.~A.~Williams$^{21}$, B.~Zitzer$^{33}$ (the VERITAS
Collaboration) \\
(Affiliations can be found after the references)}

\begin{document}

\date{Accepted 2014 October 24.  Received 2014 October 23; in original form 2014 August 6}

\maketitle

\label{firstpage}

\begin{abstract}

We report on multifrequency observations performed during 2012 December--2013
August of the first narrow-line Seyfert 1 galaxy detected in $\gamma$ rays,
PMN J0948$+$0022 ($z$ = 0.5846). A $\gamma$-ray flare was observed by the Large Area
Telescope on board {\em Fermi} during 2012 December--2013 January, reaching a
daily peak flux in the 0.1--100 GeV energy range of (155 $\pm$ 31)$\times$10$^{-8}$
ph cm$^{-2}$ s$^{-1}$ on 2013 January 1, corresponding to an apparent
isotropic luminosity of $\sim$1.5$\times$10$^{48}$ erg s$^{-1}$. The $\gamma$-ray flaring period
triggered {\em Swift} and VERITAS observations in addition to radio and
optical monitoring by OVRO, MOJAVE, and CRTS. A strong flare was observed in
optical, UV, and X-rays on 2012 December 30, quasi-simultaneously to the
$\gamma$-ray flare, reaching a record flux for this source from optical to
$\gamma$ rays. VERITAS observations at very high energy (E $>$ 100 GeV) during 2013 January
6--17 resulted in an upper limit of $F_{>0.2\,\mathrm{TeV}} < 4.0
\times10^{-12}\,\mathrm{ph}\,\mathrm{cm}^{-2}\,\mathrm{s}^{-1}$. We compared the
spectral energy distribution (SED) of the flaring state in 2013 January with
that of an intermediate state observed in 2011. The two SEDs, modelled as
synchrotron emission and an external Compton scattering of seed photons from a
dust torus, can be modelled by changing both the electron distribution parameters and the magnetic field.

\end{abstract}


\section{Introduction}

Relativistic jets are the most powerful manifestations of the release of
energy produced by super-massive black holes (SMBH) in the centers of active galactic nuclei (AGN). In about 15 per cent of AGN the accretion disc is at the base of a bipolar outflow of relativistic plasma, which may extend well beyond the
host galaxy, forming the spectacular lobes of plasma visible in radio. The jet
emission is observed across the entire electromagnetic spectrum. When the jet
axis is closely aligned with our line of sight, the rest-frame radiation is
strongly amplified due to Doppler boosting with a large fraction of the output
observed at high energy, giving rise to the blazar phenomenon. Radio galaxies
are typically viewed at larger angles, with a less extreme amplification of the emission
\citep{urry95}. As a consequence, a very small fraction of radio galaxies is
observed at high energy \citep{ackermann11}. The discovery in 2008 by the Large Area Telescope (LAT) on board the
{\em Fermi} satellite of $\gamma$-ray emission from radio-loud narrow-line
Seyfert 1s (NLSy1) revealed the presence of a possible new class of AGNs with
relativistic jets. PMN J0948$+$0022 was the first NLSy1 detected by {\em
  Fermi}-LAT \citep{abdo09a}, followed by the detection of four other NLSy1s in
$\gamma$ rays \citep{abdo09c, dammando12}.

NLSy1 is a class of AGN identified by \citet{osterbrock85} and defined in terms of the optical properties of its members: narrow permitted lines (FWHM
(H$\beta$) $<$ 2000 km s$^{-1}$), [OIII]/H$\beta$ $<$ 3, and a bump due to Fe II \citep[see e.g.][for a review]{pogge00}. Variability and spectral properties from radio to $\gamma$-ray bands, together
with the spectral energy distribution (SED) modeling, indicate a blazar-like
behaviour of these individual objects \citep[e.g.][]{abdo09c,dammando13a,foschini12}. One
of the most surprising aspects related to these NLSy1s was the detection of
flaring activity in $\gamma$ rays. A few strong
$\gamma$-ray flares were observed from PMN J0948$+$0022
\citep{donato10,foschini10, dammando11}, with an apparent isotropic
$\gamma$-ray luminosity up to 10$^{48}$ erg s$^{-1}$, comparable to that of
the bright flat spectrum radio quasars \citep[FSRQ; e.g., PKS
1510$-$089;][]{orienti13}. Similar $\gamma$-ray flares have been observed from two other NLSy1s: SBS 0846$+$513 \citep{donato11, dammando13a}, and 1H 0323+342
\citep{carpenter13}. In addition, the high apparent $\gamma$-ray luminosity observed in these objects suggests a large
Doppler boosting and thus a small viewing angle, similar to blazars. These are
important indications that NLSy1s are able to
host relativistic jets as powerful as those in blazars.

PMN J0948$+$0022 could be considered the archetypal object of a new class of $\gamma$-ray emitting AGNs; therefore
several multifrequency campaigns were performed to investigate in detail its
characteristics over the whole electromagnetic spectrum \citep{abdo09b,
  foschini11, foschini12, dammando14}. New $\gamma$-ray flaring activity
from PMN J0948$+$0022 was detected by {\em Fermi}-LAT in 2013 January
\citep{dammando13c}, triggering follow-up observations in optical, UV, and
X-rays by {\em Swift}, and for the first time at very high energy (VHE; E $>$ 100 GeV) by VERITAS. These data
are complemented by the monitoring campaigns performed by the Owens Valley Radio Observatory
(OVRO) and the Monitoring Of Jets in Active galactic nuclei with VLBA
Experiments (MOJAVE\footnote{\url{http://www.physics.purdue.edu/MOJAVE/}}) at 15
GHz, and by the Catalina Real-time Transient Survey (CRTS\footnote{{\url{http://crts.caltech.edu/}}}) in $V$-band.

The aim of the paper is to discuss the radio-to-$\gamma$-ray activity from PMN J0948$+$0022 during 2012 December--2013 August. This allows us to characterize the SED of the 2013 flaring state of the source in order to compare the SED with that observed in a different activity state observed in 2011 May, to investigate the emission mechanisms of this source and to discuss its properties in the context of the blazar scenario.  

The paper is organized as follows. In Sections 2 and 3, we report the LAT and
VERITAS data analysis and results, respectively. The X-ray, UV, and optical
data collected by {\em Swift} are presented in Section 4, together with the
optical data collected by CRTS. Radio data collected by the OVRO 40 m telescope and MOJAVE are presented in Section
5. In Section 6, we discuss the properties of the source and the modelling of the SEDs, and draw our conclusions in Section 7.

Throughout the paper the quoted uncertainties are given at the 1$\sigma$ level, unless otherwise stated, and the photon indices are
parameterized as $dN/dE \propto E^{-\Gamma_{\nu}}$, where
  $\Gamma_{\nu}$ is the photon index at the different energy bands. We adopt a
  $\Lambda$ cold dark matter ($\Lambda$--CDM) cosmology with $H_0$ = 71 km
  s$^{-1}$ Mpc$^{-1}$, $\Omega_{\Lambda} = 0.73$, and $\Omega_{\rm m} =
  0.27$. The corresponding luminosity distance at $z =0.5846$ \citep{schneider10}, the source redshift, is d$_L =  3413$\ Mpc, and 1 arcsec corresponds to a projected size of 6.59 kpc.

\section{{\em Fermi}-LAT Data: Analysis and Results}
\label{FermiData}

The {\em Fermi}-LAT  is a pair-conversion telescope operating from 20 MeV to
$>$ 300 GeV. It has a large peak effective area ($\sim$ 8000 cm$^{2}$ for 1
GeV photons), an energy resolution of typically $\sim$10 per cent, and a field
of view of about 2.4 sr with single-photon angular resolution (68 per cent containment radius) of 0\fdg6 at {\it E} = 1 GeV on-axis. Further details about the {\em Fermi}-LAT are given in \citet{atwood09}. 

The LAT data reported in this paper were collected from 2012 December 1 (MJD
56262) to 2013 August 31 (MJD 56535). During this time, the {\em Fermi}
observatory operated almost entirely in survey mode.  The analysis was
performed with the \texttt{ScienceTools} software package version
v9r32p5 \footnote{\url{http://fermi.gsfc.nasa.gov/ssc/data/analysis/software/}}. 
Only events belonging to the `Source' class were used\footnote{http://fermi.gsfc.nasa.gov/ssc/data/analysis/documentation/\\Cicerone/Cicerone\_LAT\_IRFs/IRF\_overview.html}. The time intervals when the rocking angle of the LAT was greater than 52$^{\circ}$ were rejected. In addition, a cut on the
zenith angle ($< 100^{\circ}$) was applied to reduce contamination from the
Earth limb $\gamma$ rays, which are produced by cosmic rays interacting with
the upper atmosphere. The spectral analysis was performed with the instrument
response functions \texttt{P7REP\_SOURCE\_V15} using an unbinned
maximum-likelihood method implemented  in the tool \texttt{gtlike}. Isotropic (iso\_source\_v05.txt) and Galactic diffuse emission (gll\_iem\_v05\_rev1.fit) components were used to model the background\footnote{http://fermi.gsfc.nasa.gov/ssc/data/access/lat/\\BackgroundModels.html}. The normalizations of both components were allowed to vary freely during the spectral fitting. 

We analysed a region of interest of $10^{\circ}$ radius centred at the
location of PMN J0948$+$0022. We evaluated the significance of the
$\gamma$-ray signal from the source by means of the maximum-likelihood test
statistic TS = 2 (log$L_1$ - log$L_0$), where $L$ is the likelihood of the
data given the model with ($L_1$) or without ($L_0$) a point source at the
position of PMN\,J0948$+$0022 \citep[e.g.,][]{mattox96}. The source model used
in \texttt{gtlike} includes all of the point sources from the second {\em
  Fermi}-LAT catalogue \citep[2FGL;][]{nolan12} and the preliminary list of
the third {\em Fermi}-LAT catalogue (3FGL; Ackermann et al., in prep.) that
fall within $15^{\circ}$ of the source. The spectra of these sources were
parametrized by power-law functions, except for 2FGL\,J0909.1$+$0121 and
2FGL\,J1023.6$+$0040 for which we used a log-parabola as in the 2FGL
catalogue. A first maximum-likelihood analysis was performed in the 0.1--100
GeV energy range to find and remove from the model those sources that have TS $<$ 10 and/or the predicted number of counts based on the fitted model $N_{\rm pred} < 3 $. A second maximum-likelihood analysis was performed on the updated source model. In the fitting procedure, the normalization factors and the photon indices of the sources lying within 10$^{\circ}$ of PMN\,J0948$+$0022 were left as free
parameters. For the sources located between 10$^{\circ}$ and 15$^{\circ}$ from our target, we kept the normalization and the photon index fixed to the values from the 2FGL catalogue or from the 3FGL catalogue for those sources not reported in the former catalogue.

Integrating over the period 2012 December 1 -- 2013 August 31 (MJD
56262--56535) in the 0.1--100 GeV energy range, using a log-parabola (LP), $dN/dE \propto$ $(E/E_{0})^{-\alpha-\beta \, \log(E/E_0)}$, as
in the 2FGL catalogue, the fit yielded for PMN J0948$+$0022 a TS = 638, with an average flux of
(15.5 $\pm$ 1.1) $\times$10$^{-8}$ ph cm$^{-2}$ s$^{-1}$, a spectral slope of
$\alpha$ = 2.55 $\pm$ 0.08 at the reference energy $E_0$ = 271 MeV, and a
curvature parameter $\beta$ = 0.05 $\pm$ 0.02. Using a simple power law (PL) we obtained a TS = 630, with a photon index of $\Gamma$ = 2.62 $\pm$ 0.06 and an average flux of (15.9 $\pm$ 1.0)
$\times$10$^{-8}$ ph cm$^{-2}$ s$^{-1}$. We used a likelihood ratio test (LRT)
to check a PL model
(null hypothesis) against a LP model (alternative hypothesis). These values
may be compared, following \citet{nolan12}, by defining the curvature test
statistic TS$_{\rm curve}$=(TS$_{\rm LP}$ - TS$_{\rm PL}$). The LRT results in a TS$_{\rm curve}$ = 8, corresponding to
a $\sim$2.8$\sigma$ difference; therefore a hint of curvature was observed in the $\gamma$-ray spectrum of PMN J0948$+$0022 during 2012 December--2013 August.
By considering only the period 2012 December 1 - 2013 June 14, in which the
source was regularly detected on weekly time-scales, the maximum-likelihood
analysis results for the LP and PL models in a TS$_{\rm LP}$ = 677 and
TS$_{\rm PL}$ = 660, respectively. The corresponding curvature test statistic
is TS$_{\rm curve}$= 17, which corresponds to a  significance of
$\sim$4.3$\sigma$. This indicates that the LP is preferred over a simple PL
for this time interval, in agreement with the results obtained for the observations during 2008 August--2010 August \citep{nolan12}.

\begin{figure}
\centering
\includegraphics[width=8.0cm]{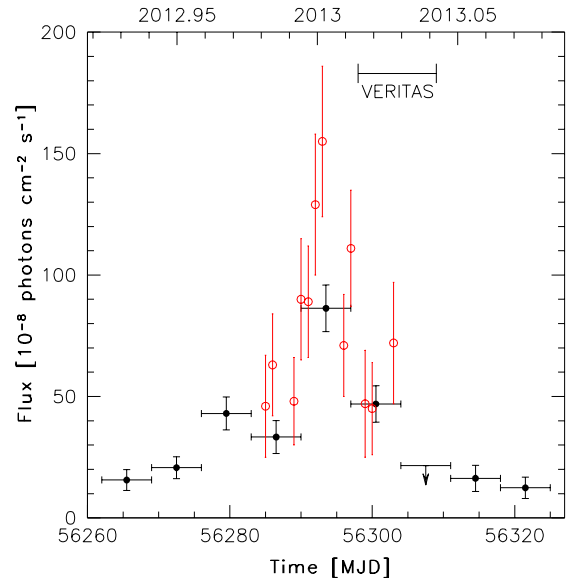}
\caption{Integrated flux light curve of PMN J0948$+$0022 obtained by {\em Fermi}-LAT in the 0.1--100 GeV energy range during 2012 December 1--2013
  January 31 with 7-day time bins. Arrow refers to 2$\sigma$ upper limits on
  the source flux. Upper limits are computed when $TS$ $<$ 10. Open circles
  represent daily fluxes reported for the periods of high activity. No upper limits are shown for daily data. The horizontal line indicates the period of the VERITAS observation.}
\label{LAT}
\end{figure}

Figure \ref{LAT} shows the $\gamma$-ray light curve for the period 2012
December 1--2013 January 31 using a LP model and 1-week time bins. For the highest
significance periods we also reported fluxes in 1-day time intervals. For each
time bin, the spectral parameters for PMN\,J0948$+$0022 and for all the
sources within 10$^{\circ}$ from the target were frozen to
the value resulting from the likelihood analysis over the whole period. If TS
$<$ 10, 2$\sigma$ upper limits were calculated. All uncertainties in measured
$\gamma$-ray flux reported throughout this paper are statistical
only. The systematic uncertainty on the effective area amounts to 10 per cent below 100 MeV, decreasing
linearly with the logarithm of energy to 5 per cent between 316 MeV and 10 GeV, and increasing linearly with the logarithm
of energy up to 15 per cent at 1 TeV \citep{ackermann12b}.

During the period of high activity (2012 December 15 - 2013 January 11; MJD 56276--56303),  the
fit with a LP model in the 0.1--100 GeV energy range results in a TS
= 410, with an average flux of (46.8 $\pm$ 4.7)$\times$10$^{-8}$ ph cm$^{-2}$ s$^{-1}$, a spectral slope of $\alpha$ = 2.37 $\pm$ 0.10 at the
reference energy $E_0$ = 271 MeV, and a curvature parameter $\beta$ = 0.10
$\pm$ 0.04. This suggests marginal spectral variability during the
$\gamma$-ray flaring activity with respect to the whole period. The emission peak was observed on 2013 January 1 (MJD 56293), with an
average flux for that day of (155 $\pm$ 31)$\times$10$^{-8}$ ph cm$^{-2}$
s$^{-1}$ in the 0.1--100 GeV energy range, corresponding to an
apparent isotropic $\gamma$-ray luminosity of $\sim$1.5$\times$10$^{48}$ erg
s$^{-1}$. This is the highest flux observed from PMN J0948$+$0022 so far in $\gamma$ rays \citep[see][]{foschini12}.

By means of the \texttt{gtsrcprob} tool and using the source model
  optimized over the period 2012 December -- 2013 August, we estimated that the highest energy
photon detected from PMN J0948$+$0022 was observed on 2013 January 5 at a
distance of 0\fdg06 from the source with a probability of 98.5\% to be associated with the target and an energy of 7.2 GeV. Analyzing the
LAT data collected over the whole period in the 1-100 GeV energy range with a
simple power law, the fit yielded a TS = 143, with a photon index of
$\Gamma_{\gamma}$ = 2.64 $\pm$ 0.19 and an average flux of (3.4 $\pm$ 0.5)
$\times$10$^{-9}$ ph cm$^{-2}$ s$^{-1}$. On the contrary, the source is not
detected at E $>$ 10 GeV (TS=1) with a 2$\sigma$ upper limit of
1.5$\times$10$^{-10}$ ph cm$^{-2}$ s$^{-1}$ (assuming a photon index $\Gamma_{\gamma}$ = 3).
 
\begin{table*}
\centering
\caption{VERITAS observations and results of PMN J0948$+$0022.}
\begin{tabular}{cccccc}
\hline
Date  & Date  & Exposure & Significance & $N_\mathrm{ex}$ & $F_{>0.2\,\mathrm{TeV}}$  \\
(MJD) & (UT) & (min) & ($\sigma$) & & ($10^{-11}\,\mathrm{ph}\,\mathrm{cm}^{-2}\,\mathrm{s}^{-1}$) \\
\hline
56298  & 2013-01-06     &33&	-1.2&	-9.8&	$<0.84$\\
56299  & 2013-01-07	&77&	-0.3&	-3.3&	$<0.54$\\
56301  & 2013-01-09	&75&	 1.1&	15.0&	$<1.12$\\
56305  & 2013-01-13	&25&	 0.0&	-0.1&	$<1.22$\\
56306  & 2013-01-14	&17&	 0.0&	 0.0&	$<1.53$\\
56307  & 2013-01-15	&13&	 1.9&	10.6&	$<4.58$\\
56308  & 2013-01-16	&50&	 0.3&	 3.0&	$<1.08$\\
56309  & 2013-01-17	&25&	-1.2&	-8.8&	$<2.22$\\
\hline
all data & -            & 315 & 0.4 & 10.6 & $<0.40$\\
\hline
\label{ver}
\end{tabular}
\end{table*}

\begin{table*}
\caption{Log and fitting results of {\em Swift}-XRT observations of
  PMN\,J0948$+$0022 using a power-law model with a HI column density fixed to the
  Galactic value in the direction of the source. $^{a}$Unabsorbed flux.}
\begin{center}
\begin{tabular}{ccccc}
\hline 
\multicolumn{1}{c}{Date} &
\multicolumn{1}{c}{Date} &
\multicolumn{1}{c}{Net exposure time} &
\multicolumn{1}{c}{Photon index} &
\multicolumn{1}{c}{Flux 0.3--10 keV$^{a}$} \\
\multicolumn{1}{c}{(MJD)} &
\multicolumn{1}{c}{(UT)} &
\multicolumn{1}{c}{(s)} &
\multicolumn{1}{c}{($\Gamma_{\rm X}$)} &
\multicolumn{1}{c}{(10$^{-12}$ erg cm$^{-2}$ s$^{-1}$)} \\
\hline
56264 & 2012-12-03 & 1963 & 1.53 $\pm$ 0.25 & $3.3 \pm 0.5$ \\
56286 & 2012-12-25 & 1893 & 1.51 $\pm$ 0.21 & $5.1 \pm 0.5$ \\
56291 & 2012-12-30 &  722 & 1.77 $\pm$ 0.16 & $13.6 \pm 1.3$ \\
56295 & 2013-01-03 & 2972 & 1.48 $\pm$ 0.15 & $6.8 \pm 0.5$ \\
56303 & 2013-01-11 & 2932 & 1.60 $\pm$ 0.17 & $6.9 \pm 0.6$ \\
56309 & 2013-01-17 & 3284 & 1.67 $\pm$ 0.10 & $10.6 \pm 0.6$ \\
\hline
\end{tabular}
\end{center}
\label{XRT}
\end{table*}

\begin{table*}
\caption{Results of the {\em Swift}-UVOT observations of PMN J0948$+$0022 in magnitudes.}
\begin{center}
\begin{tabular}{cccccccc}
\hline 
\multicolumn{1}{c}{Date (MJD)} &
\multicolumn{1}{c}{Date (UT)} &
\multicolumn{1}{c}{$v$} &
\multicolumn{1}{c}{$b$} &
\multicolumn{1}{c}{$u$} &
\multicolumn{1}{c}{$w1$} &
\multicolumn{1}{c}{$m2$} &
\multicolumn{1}{c}{$w2$} \\
\hline
56264 & 2012-12-03 & -- & -- & 17.43$\pm$0.03 & -- & -- & -- \\
56286 & 2012-12-25 &  18.11$\pm$0.18 & 18.29$\pm$0.10 & 17.34$\pm$0.07 & 17.29$\pm$0.07 & 17.46$\pm$0.05 & 17.46$\pm$0.04 \\
56291 & 2012-12-30 &  16.01$\pm$0.10 & 16.59$\pm$0.07 & 15.86$\pm$0.06 & 16.20$\pm$0.04 & 15.88$\pm$0.08 & 16.08$\pm$0.05 \\
56295 & 2013-01-03 &  17.48$\pm$0.09 & 18.00$\pm$0.07 & 17.19$\pm$0.06 & 17.07$\pm$0.05 & 17.16$\pm$0.06 & 17.12$\pm$0.04 \\
56303 & 2013-01-11 &  17.70$\pm$0.11 & 17.85$\pm$0.06 & 17.02$\pm$0.05 & 16.94$\pm$0.05 & 17.01$\pm$0.05 & 17.14$\pm$0.04 \\
56309 & 2013-01-17 &  17.45$\pm$0.09 & 17.81$\pm$0.06 & 16.99$\pm$0.05 & 17.00$\pm$0.05 & 17.02$\pm$0.05 & 17.04$\pm$0.04 \\
\hline
\end{tabular}
\end{center}
\label{uvot}
\end{table*}

\section{VERITAS data: Analysis and Results}

The VERITAS observatory \citep{VERITAST1,VERITAS} is an array of four
imaging atmospheric Cherenkov telescopes (IACT) located at the Fred Lawrence Whipple Observatory near 
Tucson, Arizona. Each telescope consists of a 12-m diameter reflector and a photomultiplier camera 
covering a field of view of 3\fdg5.
The array has an effective area of $\sim 5 \times 10^4\,\mathrm{m}^{2}$ between 0.2 and 
10\,TeV.  

VERITAS can detect a source with 2-3 per cent of the Crab Nebula 
flux\footnote{$1\,\mathrm{Crab}=2.1\times10^{-10}\,\mathrm{ph}\,\mathrm{cm}^{-2}\,\mathrm{s}^{-1}$ for E $>$ 200 GeV \citep[][]{hillas-crab}} in 300 min of 
exposure, 
with improved sensitivity to soft-spectrum sources following an update of the camera 
photomultipliers in Summer 2012 \citep{kieda}.
The angular and energy resolution for reconstructed $\gamma$-ray showers are 0\fdg1 and 
15 per cent, respectively, at 1\,TeV.

After a trigger from {\em Fermi}-LAT \citep{dammando13c}, VERITAS observations of PMN J0948$+$0022 took place 
between 2013 January 6 (MJD 56298) and January 17 (MJD 56309), covering the decay 
of the $\gamma$-ray flare. Observations consisted of 30 min exposures pointing 0\fdg5, offset in \textit{wobble} observing
mode \citep{fomin} at a median zenith angle of $32^{\circ}$. About one third of the collected data 
had to be rejected because of cloud coverage, leaving an effective dead-time corrected exposure of 
315 min.

A description of the VERITAS analysis can be found in \citet{ver_analysis2} and \citet{ver_analysis}. To ensure a robust estimation of the shower
parameters, at least two telescope images with a total signal size equivalent to $>74$\ photoelectrons were required for an event to be reconstructed. Background-rejection cuts, optimized for a moderately bright soft-spectrum source, were applied to 
remove over 99.9 per cent of the charged cosmic rays. Events with
reconstructed arrival direction $\theta <$ 0\fdg1 to the position of PMN
J0948$+$0022 were selected as signal. The number of background events in the
signal region was estimated from off-source regions in the same field of view
using the \textit{reflected region} technique \citep{refl}.

The analysis revealed no significant $\gamma$-ray signal from PMN
J0948$+$0022. An excess ($N_{\rm ex}$) of 11 candidate $\gamma$-ray events
over a background of 615 was measured, corresponding to a
significance of 0.4 standard deviations according to Eq.~17 in \citet{lima},
with $N_{\mathrm{on}}=626$, $N_{\mathrm{off}}=5211$, and a ratio of the
on-source time to the off-source time $\alpha=0.118$. 
Assuming a 
power-law spectrum with photon index $\Gamma_{\gamma} = 3.5$, a 99 per cent confidence level (c.l.) flux 
upper limit of 
$F_{>0.2\,\mathrm{TeV}} < 4.0 \times 10^{-12}\,\mathrm{ph}\,\mathrm{cm}^{-2}\,\mathrm{s}^{-1}$ was derived using the 
prescription by \citet{rolke}. The systematic uncertainty in the
flux measurement is approximately 40 per cent.
Differential upper limits (95 per cent c.l., $\Gamma_{\gamma} = 3.5$ to match the LAT spectrum during the period of high activity) at the energy threshold of the analysis (170 GeV) and where the
correlation between flux level and spectral shape is minimal (300 GeV) are shown in Fig.~\ref{SED_extrapolated}.
Compatible results were obtained using an independent analysis package with a different set of
event-selection cuts.

Given the variable nature of the $\gamma$-ray emission from PMN J0948$+$0022,
a search for signal in nightly bins (Table~\ref{ver}) as well as in 30-min runs was also conducted. However, no 
significant $\gamma$-ray excess was found at those time-scales either.  

\section{X-ray to optical data}

\subsection{{\em Swift} data analysis}
\label{SwiftData}

The {\em Swift} satellite \citep{gehrels04} performed six observations
of PMN\,J0948$+$0022 between 2012 December and 2013 January. The observations were
carried out with all three instruments on board: the X-ray Telescope
\citep[XRT;][0.2--10.0 keV]{burrows05}, the Ultraviolet/Optical Telescope
\citep[UVOT;][170--600 nm]{roming05} and the Burst Alert Telescope
\citep[BAT;][15--150 keV]{barthelmy05}. The hard X-ray flux of this source is below the sensitivity of the BAT
instrument for the short exposures of these observations, therefore the data from this instrument are not used.
Moreover, the source was not present in the {\em Swift} BAT 70-month hard X-ray catalogue \citep{baumgartner13}.

The XRT data were processed with standard procedures (\texttt{xrtpipeline
  v0.12.8}), filtering, and screening criteria using the \texttt{HEAsoft}
package (v6.13). The data were collected in photon counting mode for all of the observations. The source count rate
was low ($<$ 0.5 counts s$^{-1}$); thus pile-up correction was not
required. Source events were extracted from a circular region with a radius of
20 pixels (1 pixel = 2.36 arcsec), while background events were extracted from a circular region with radius of 50 pixels away from the source
region and from other bright sources. Ancillary response files were generated with \texttt{xrtmkarf}, and
account for different extraction regions, vignetting and point-spread function
corrections. We used the spectral redistribution matrices v013 in the
Calibration data base maintained by
HEASARC\footnote{\url{http://heasarc.nasa.gov/}}. The spectra with low numbers of
photons collected ($<$ 200 counts) were rebinned with a minimum of 1 count per
bin and the Cash statistic \citep{cash79} was used. We fitted the spectra with an absorbed power-law using the photoelectric absorption model
\texttt{tbabs} \citep{wilms00}, with a neutral hydrogen column density fixed
to its Galactic value \citep[5.07$\times$10$^{20}$cm$^{-2}$;][]{kalberla05}.
The fit results are reported in Table~\ref{XRT}, and the 0.3--10 keV fluxes are
shown in Fig.~\ref{MWL}. 

UVOT data in the $v$, $b$, $u$, $w1$, $m2$, and $w2$ filters were reduced with the \texttt{HEAsoft} package v6.13 and the 20120606 CALDB-UVOTA
release. We extracted the source counts from a circle with 5 arcsec radius
centred on the source and the background counts from a circle with 10 arcsec
radius in a nearby source-free region. The observed magnitudes are reported in Table~\ref{uvot}.
As in, e.g., \citet{raiteri11}, we calculated the effective wavelengths,
count-to-flux conversion factors ($\rm CF_\Lambda$), and amount of Galactic extinction in the UVOT
bands ($\rm A_\Lambda$) by convolving the mean Galactic extinction law by \citet{cardelli89} with a power-law fit to the source flux
and with the filter effective areas. To obtain the UVOT de-reddened fluxes 
we multiplied the count rates for the $\rm CF_\Lambda$ and corrected for the
corresponding Galactic extinction values $\rm A_\Lambda$. The UVOT density fluxes are reported in Fig~\ref{MWL}. 

\subsection{CRTS data}

\begin{figure*}
\centering
\includegraphics[width=13cm]{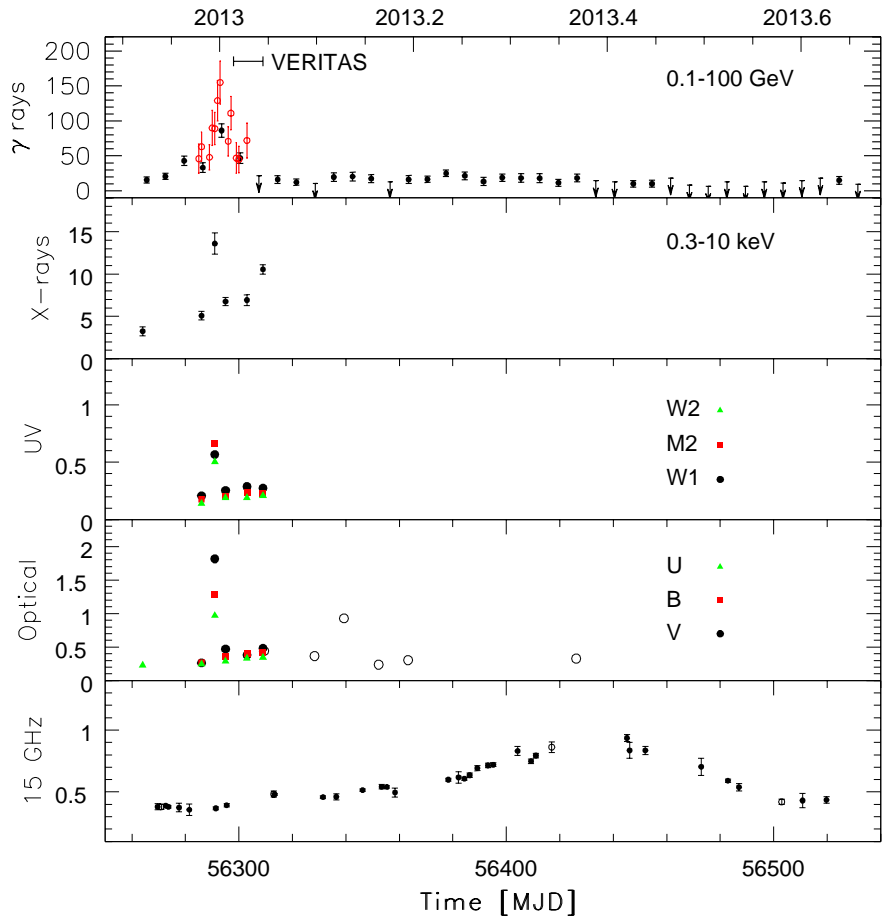}
\caption{Multifrequency light curve for PMN J0948$+$0022. The period covered
  is 2012 December 1 -- 2013 August 31 (MJD 56262--56535). The data were
  collected (from top to bottom) by {\em Fermi}-LAT ($\gamma$ rays; in units
  of 10$^{-8}$ ph cm$^{-2}$ s$^{-1}$), {\em Swift}-XRT (0.3--10 keV; in units
  of 10$^{-12}$ erg cm$^{-2}$ s$^{-1}$), {\em Swift}-UVOT ($w1$, $m2$, and
  $w2$ filters; in units of mJy), CRTS (open circles; in units of mJy) and {\em Swift}-UVOT ($v$, $b$, and $u$ filters; in units of
  mJy), MOJAVE (open circles) and OVRO (15 GHz; in units of Jy). The horizontal line in the top panel indicates the period of the VERITAS observation.}
\label{MWL}
\end{figure*}

The source has been monitored by the CRTS \citep{drake09, djorgovski11}, using
the 0.68-m Schmidt telescope at Catalina Station, AZ, and an unfiltered
CCD. The typical cadence is four exposures separated by 10 min in a given
night; this may be repeated up to four times per lunation, over a period of
$\sim$6--7 months each year, while the field is observable.  Photometry is
obtained using the standard Source-Extractor package \citep{bertin96}, and
transformed from the unfiltered instrumental magnitude to Cousins $V$\footnote{\url{http://nesssi.cacr.caltech.edu/DataRelease/FAQ2.html\#improve}} by
$V$ = $V_{\rm CSS}$ + 0.31($B-V$)$^{2}$ + 0.04. We averaged the values obtained during the same observing night. The flux densities collected by CRTS in $V$ band are reported in Fig.~\ref{MWL}.

\section{radio data}

\subsection{OVRO data analysis}

As part of an ongoing blazar monitoring program, the OVRO 40-m radio telescope
has observed PMN J0948$+$0022 at 15~GHz regularly since the end of 2007
\citep{richards11}. This monitoring program includes about 1700 known and
likely $\gamma$-ray loud blazars above declination $-20^{\circ}$. The sources in this programme are observed in total intensity twice per week with a 4~mJy
(minimum) and 3 per cent (typical) uncertainty on the flux densities. Observations were performed
with a dual-beam (each 2.5~arcmin FWHM) Dicke-switched system using cold sky in the off-source beam as the reference. Additionally, the
source is switched between beams to reduce atmospheric variations. The absolute flux density scale is calibrated using observations of
3C~286, adopting the flux density (3.44~Jy) from \citet{baars77}. This results in about a 5 per cent absolute scale uncertainty, which is not reflected in the plotted errors. 

\subsection{MOJAVE data analysis}\label{MOJAVE}

We investigated the parsec-scale morphology and flux density variability at 15
GHz by means of 6-epoch Very Long Baseline Array data from the {\small MOJAVE} programme
\citep{lister09}. The data sets span the time interval between 2012 July and 2013 July, in order to overlap with the {\em Fermi}--LAT data. We imported the calibrated {\it uv} data into the National Radio Astronomy Observatory
{\small AIPS} package. In addition to the total intensity images, we produced
the Stokes’ Q and U images, to derive information on the polarized
emission. The flux density was derived by means of the {\small AIPS} task
JMFIT which performs a Gaussian fit on the image plane. Total intensity flux
density and polarization information are reported in Table \ref{tab_moj}. The
uncertainties on the flux density scale are less than 5 per cent, and the errors
on the polarization angle are $\sim$5\degr \citep{lister13}.

\begin{table}
\caption{Flux density and polarization of PMN\,J0948+0022 from 15 GHz MOJAVE data.}
\begin{center}
\begin{tabular}{cccccc}
\hline
Date  & Date & $S_{\rm Core}$$^{a}$&$S_{\rm Jet}$$^{a}$&$S_{\rm pol}$$^{a}$&EVPA$^{b}$ \\
(MJD) & (UT) &  (mJy) &  (mJy)  & (mJy) (\%)& (deg) \\
\hline
56120 & 2012-07-12 & 325 & 5 & 3 \,\, (0.9\%)& 96\\
56242 & 2012-11-11 & 336 & 5 & 3 \,\, (0.9\%)& 86\\
56271 & 2012-12-10 & 380 & 6 & 4 \,\, (1.0\%)& 97\\
56313 & 2013-01-21 & 483 & 4 & 3 \,\, (0.6\%)& 130\\
56417 & 2013-05-05 & 862 & 5 & 6 \,\, (0.7\%)& 96\\
56503 & 2013-07-30 & 420 & 3 & 12 \,\, (2.9\%)& 90\\
\hline
\tablenotetext{a}{Uncertainties on the flux densities are less than 5 per cent.}
\tablenotetext{b}{Uncertainties on the polarization angles are 5\degr.}
\end{tabular}
\label{tab_moj}
\end{center}
\end{table}	

\section{Discussion} 

\subsection{Radio to $\gamma$-ray behaviour}

During 2012 December -- 2013 January the NLSy1 PMN J0948$+$0022 showed new
flaring activity in $\gamma$ rays, triggering follow-up observations by {\em
  Swift} and VERITAS.   In Fig.~\ref{MWL}, we compare the $\gamma$-ray light curve collected by {\em
  Fermi}-LAT in the 0.1--100 GeV energy range with the X-ray (0.3--10 keV), UV
($w1$, $m2$, and $w2$ filters), optical ($v$, $b$, and $u$ filters), and radio
(15 GHz) light curves collected by {\em Swift} (XRT and UVOT), CRTS, MOJAVE, and OVRO.

The $\gamma$-ray flaring period peaked on 2013 January 1 (MJD 56293) with
a variability amplitude (calculated as the ratio between the maximum and
minimum flux) of $\sim$15 between 2012 December and 2013 August, and flux
variations on a daily time-scale. The peak flux reported here,
(155 $\pm$ 31)$\times$10$^{-8}$ ph cm$^{-2}$ s$^{-1}$, is the greatest detected from this source so far at $\gamma$ rays \citep[see e.g.,][]{foschini12}. The $\gamma$-ray flaring activity ended in 2013 mid-January, and no further significant increase of the $\gamma$-ray flux was
observed up to 2013 August. Only marginal spectral variability was observed
during this flaring activity, similar to the behaviour shown in \citet{foschini12} for the period 2008 August -- 2011 December.

During the highest $\gamma$-ray activity a quasi-simultaneous increase was
observed from optical to X-rays. {\em Swift}/XRT observed the source with a
0.3--10 keV flux in the range (3.3--13.6)$\times$10$^{-12}$ erg cm$^{-2}$
s$^{-1}$,  with a variability amplitude of $\sim$4 (Table~\ref{XRT}). 
The peak of the X-ray flux was observed on 2012 December 30 (MJD 56291), when
PMN J0948$+$0022 reached the highest flux detected for this source so far
\citep[for a comparison, see][]{foschini12,dammando14}. No significant
spectral change in the X-ray band was observed during the {\em Swift} observations, with the photon index ranging between 1.5
and 1.8, similarly to what was observed in 2011 \citep{dammando14}. It is worth mentioning that a second increase of the X-ray flux was
observed on 2013 January 17 (MJD 56309), but no corresponding increase of
activity was observed in $\gamma$-ray, UV, and optical bands. 

The optical and UV emission significantly increased over only 5 days, from 2012
December 25 (MJD 56286) to December 30 (MJD 56291), suggesting that the
dominant contribution to the continuum flux should come from the synchrotron
emission. However, the variability amplitude decreases with frequency: from
$\sim$7 in $V$-band to $\sim$3.5 in the UV part of the spectrum. This could be
due to the contribution in UV of the thermal emission from the accretion disc,
which is well identified during low activity periods of this source
\citep{foschini12}. A similar behaviour was already noticed in some FSRQs
\citep[e.g. 3C 454.3,][]{raiteri11}. After the $\gamma$-ray flare, two
observations were performed in $V$-band by the Steward Observatory as part of
the blazar monitoring programme of the University of
Arizona\footnote{\url{http://james.as.arizona.edu/~psmith/Fermi/}} Optical polarization of 4.0$\pm$0.1 per cent and 1.1$\pm$0.1 per cent was estimated on 2013
January 17 and 18, respectively, while the position angle on the sky of the
linear polarization changed from 100\degr $\pm$ 1\degr\, to 26\degr $\pm$ 3\degr\,. The value observed on
2013 January 17 is not as high as the 12.3 per cent observed for this
source on 2013 May 25 \citep{eggen13}, but the change of a factor of $\sim$4
in one day is notable as well as the significant change of the observed polarization angle. The rapid variation of optical polarization is a characteristic associated
with the variation of synchrotron emission in blazars. The variation of both
the fractional polarization and polarization angle may be an indication of a turbulent magnetic field \citep[e.g.,][]{marscher14}.

PMN J0948$+$0022 was highly variable at 15 GHz during the OVRO 40 m telescope
monitoring, with the flux density varying from 357 mJy (on MJD
56281; 2012 December 20) to 936 mJy (on MJD 56445; 2013 June 2), as shown in Fig.~\ref{MWL}. The source was observed in a low state at 15 GHz during the peak of the $\gamma$-ray activity. On the other hand, no significant $\gamma$-ray activity was detected at the time of the peak of the radio activity. 
If the $\gamma$-ray and radio activity are connected, the observed peak at 15
GHz is produced by the same  disturbance of the jet responsible for the
$\gamma$-ray flare delayed by $\sim$5 months. This delay may be explained
  by synchrotron self-absorption opacity effects in combination with shock
  propagation effects. As an alternative, the $\gamma$-ray and radio emission
  may originate in different parts of a bending
and inhomogeneous jet, with variable orientations with respect to
  the line of sight. Therefore we may have different Doppler boostings of the zone
  responsible for the optical-to-$\gamma$-ray emission and that responsible
  for the radio emission in the two flaring periods \citep[e.g.,][]{raiteri12}. A delay of $\sim$50 days of the
15 GHz flux density peak with respect to the $\gamma$-ray one was observed
during the 2010 $\gamma$-ray flaring activity of PMN J0948$+$0022 \citep{foschini11}. On the contrary, no
obvious connection between $\gamma$ and radio emission was observed in 2011
\citep{dammando14}. The lack of a coherent pattern suggests a complex connection between the radio and
$\gamma$-ray emission for PMN J0948$+$0022, as also observed in the other
NLSy1 SBS 0846$+$513 \citep{dammando13a,dammando12} as well as in blazars
\citep[e.g. PKS 1510$-$089,][]{orienti13}.

At pc-scales the radio emission is dominated by the core component, which
accounts for $\sim$98 per cent of the total emission. A hint of the jet
emerges from the core with a position angle $\sim$30\degr, consistent with
what was found in previous works \citep[e.g.,][]{dammando14,giroletti11}. From
the pc-scale resolution images we can separate the core flux density from
the jet emission, in order to study the light curve and polarization
properties of each of them. In accordance with the OVRO monitoring, the flux
density of the core observed by MOJAVE at 15 GHz increased from 2012 July 12
to 2013 May 5, when a value of 862 mJy was reached (Table~\ref{tab_moj}). No
significant increase of the flux density from the jet was observed. A higher
polarized emission of the core ($S_{\rm pol}$) and polarization percentage were
observed on 2013 July 30, after the peak of the radio emission. The electric
vector position angle (EVPA) of the core shows a moderate change during
the MOJAVE observation, ranging from 86\degr\, to 130\degr, similar to what
was observed for the same object in 2011 \citep{dammando14}. In particular, we
noted a change of $\sim$25\degr\ between 2012 December and 2013 January, after the $\gamma$-ray flare observed by LAT.

\begin{figure}
\vspace{1.2mm} 
\includegraphics[width=8cm]{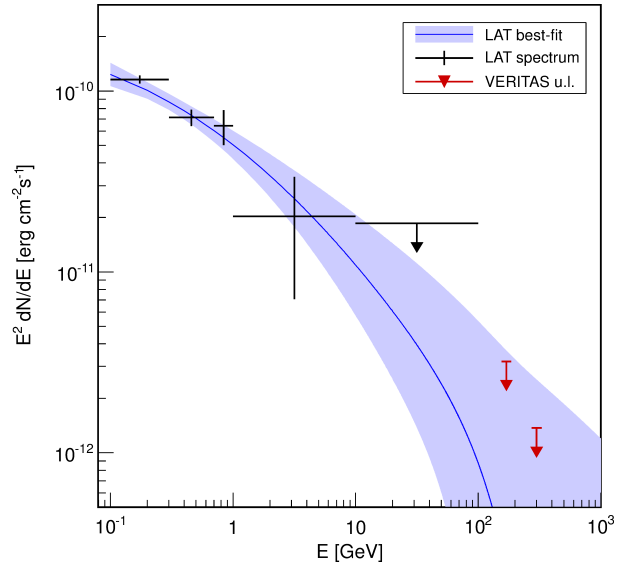}
\caption{SED of PMN J0948$+$0022 in the MeV-to-TeV energy range. The {\em
    Fermi}-LAT spectrum was extrapolated to the TeV energies and corrected for
  EBL absorption using the model of \citet{finke10}. {\em Fermi}-LAT and
  VERITAS data points and upper limits are shown. VERITAS upper limits
    are calculated at the energy threshold of the analysis  and where the
correlation between flux level and spectral shape is minimal.}
\label{SED_extrapolated}
\vspace{1.2mm}
\end{figure}

Because $\gamma$-ray emission from NLSy1s was not expected until its discovery with the {\em Fermi} satellite, no dedicated observations of these
sources were carried out at VHE before that time. The detection of a few of
these objects in $\gamma$ rays by the LAT has given rise to a new interest in their emission processes at the higher energies. 
The only $\gamma$-ray emitting NLSy1 with VHE observations in the literature
  is 1H 0323$+$342 ($z$ = 0.062), which was selected from a list of candidate
  TeV-emitting FSRQs \citep{perlman00}. It was observed by the Whipple
  telescope during 2001--2003, resulting in an upper limit of
  5.2$\times$10$^{-12}$ ph cm$^{-2}$ s$^{-1}$ above 400 GeV, with marginal
  evidence of a rate increase in the 2001--2002 light curve
  \citep{falcone04}. This paper provides the first report of an observation of
  a NLSy1 with the current generation of IACTs (i.e. VERITAS, MAGIC, and H.E.S.S.). Unfortunately the source was not detected by VERITAS at VHE in 2013 January, with an upper limit of 4.0$\times$10$^{-12}$ ph cm$^{-2}$ s$^{-1}$ above 200 GeV. This could be due to several reasons. First of all, the distance
of the source is relatively large ($z$ = 0.5846), and so most of the GeV/TeV
emission should be absorbed due to pair production from $\gamma$-ray
photons of the source and the infrared photons from the extragalactic
background light (EBL), although the most distant FSRQ detected at VHE, 3C 279
\citep{albert08}, is at a comparable distance of $z$ = 0.5362. Considering the
\citet{finke10} model for the EBL the optical depth is $\tau$ $\sim$1 for 170 GeV photons
at $z$ = 0.5846. As a comparison, using the \citet{dominguez11} model the
optical depth is $\tau$ $\sim$1 for 200 GeV photons at the same distance.
However, the VERITAS observations were carried out a few days after the peak of the $\gamma$-ray
activity from PMN J0948$+$0022 detected by LAT, thus covering only the last
part of the MeV--GeV flare. 
An extrapolation to the VHE band of the LAT measured energy spectrum averaged
over the active phase of the flare (2012 December 29 -- 2013 January 18) falls
below the VERITAS sensitivity in 5h of observation once the EBL absorption is
taken into account (Fig.~\ref{SED_extrapolated}). Assuming no additional spectral curvature than that already seen in the LAT energy range (LP with $\beta=0.1$) and the flux suppression induced by the EBL, detectable VHE emission from PMN J0948$+$0022 would be expected during GeV flares with F$_{\rm 0.1-100\,GeV}$ $\geq$ 2$\times$10$^{-6}$ ph cm$^{-2}$ s$^{-1}$.
Given the similarities between PMN J0948$+$0022 and $\gamma$-ray detected
FSRQs, the presence of a bright broad-line region (BLR) could produce a spectral break due to pair production, suppressing the flux beyond a few GeV and preventing a VHE detection. However, the role of BLR photons in the emission/absorption of VHE $\gamma$-rays is far from understood, and the detections of the FSRQ 3C 279 \citep{albert08}, 4C +21.35 \citep{aleksic11}, and PKS
1510-089 \citep{abramowski13} have shown that the spectrum of some FSRQs
extends to VHE energies during certain flaring periods. This may be an
indication that during those high activity periods the $\gamma$-ray emission was
produced outside the BLR.

\subsection{SED Modelling}\label{SED_mod}

We created two SEDs for comparing the 2013 January $\gamma$-ray flaring activity (presented in this paper) to the 2011 May intermediate state \citep[presented in][]{dammando14}. The 2013 flaring SED includes the
LAT spectrum built with data centred on 2012 December 29--2013 January 18
(MJD 56290--56310), optical, UV, and X-ray data collected by {\em Swift} (XRT
and UVOT)  on 2012 December 30 (MJD 56291), and radio data collected by OVRO
at 15 GHz on 2013 January 3 (MJD 56295). The 2011 intermediate SED was built
with LAT data centred on 2011 May 22--June 11 (MJD 55703--55723), optical, UV, and X-ray data collected by XMM-{\em Newton} (EPIC and Optical Monitor) on 2011 May 28--29 (MJD 55709--55710), and radio data collected by Effelsberg at 15 GHz and 32 GHz on 2011 May 24 (MJD 55705).

The location along the jet of the $\gamma$-ray emitting region is controversial; see the discussion by \citet{finke13_review}.  Rapid
variability \citep{tavecchio10,brown13,saito13} indicates a compact emitting region that, if the emitting region takes up the entire cross
section of the jet, argues for emission from inside the BLR. Absorption features in $\gamma$-ray spectra could also be an indication of this \citep{poutanen10,stern11,stern14}. On the other hand, $\gamma$-ray spectra that extend to VHEs without absorption features in some objects \citep{aleksic11,aleksic14,pacciani14} are a strong indication that emission takes place outside the BLR, where the dust torus is likely the dominant external radiation field.  This is corroborated by the association of $\gamma$-ray flares with the radio outbursts and the ejection of superluminal components from cores of blazars as observed with very long baseline interferometry, giving distances $\ga$ a few pc based on light travel time arguments
\citep[e.g.,][]{marscher10,orienti13}.  There is some indication that different flares, even from the same source, can occur in different
locations \citep{brown13,nalewajko14}.  Equipartition arguments can also be
invoked for locating the emitting region \citep{dermer14}. In the following, we first
model the emitting region consistent with a dust torus external radiation
field, then discuss modelling consistent with an emitting region inside the
BLR.

We modelled the SED of the two activity states of PMN J0948$+$0022 with a combination of synchrotron, synchrotron self-Compton (SSC), and external Compton (EC) non-thermal emission. The synchrotron component considered is self-absorbed below $10^{11}$\ Hz and thus cannot reproduce the radio emission. This emission is likely from the superposition of multiple self-absorbed jet components \citep{konigl81}.  We also included thermal emission by an accretion disc and dust torus.  The modelling details can be found in \citet{finke08_SSC} and
\citet{dermer09_EC}. Additionally, a soft excess was observed in the X-ray
spectrum during the 2011 state, whose origin is still under debate \citep{dammando14}.
In order to account for the soft X-ray excess in the 2011 SED, we included emission from the disc
which is Compton scattered by an optically thin thermal plasma near the
accretion disc (i.e., a corona). This was done using the routine ``{\tt SIMPL}'' \citep{steiner09}.  This routine has two free parameters: the fraction
of disc photons scattered by the corona ($f_{sc}$), and the power-law photon
index of the scattered coronal emission ($\Gamma_{sc}$). The mass of the BH
was chosen to be the same as the one used by \citet{foschini11}, M$_{\rm BH}$ =
$1.5\times10^8$ M$_\odot$, consistent with estimates by \citet{zhou03} and \citet{abdo09a}.

\begin{figure}
\vspace{1.2mm} 
\includegraphics[width=8cm]{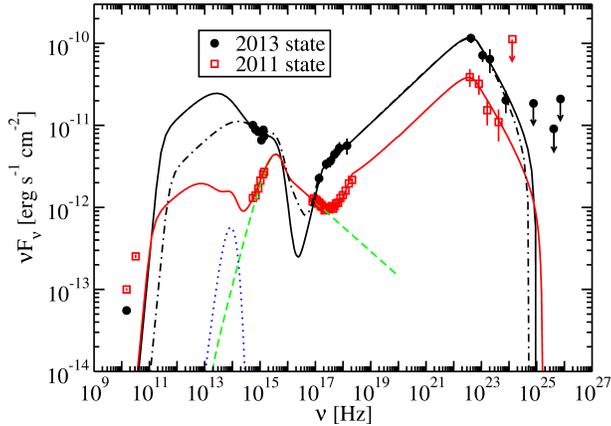}
\caption{SEDs and models for the 2013 and 2011 activity states from PMN J0948$+$0022.  The filled
circles are the data from the 2013 flaring state, and the open squares are the
data from the 2011 intermediate state taken from \citet{dammando14}. The dashed curve indicates the disc and
coronal emission, and the dotted line indicates the thermal dust
emission. Solid lines represent models consistent with scattering dust torus radiation, while the dashed-dotted curve
represents a model consistent with the scattering of BLR H$\alpha$ (6563 \AA) radiation.
Arrows refer to 2$\sigma$ upper limits on the source flux. The
VERITAS upper limits are corrected for EBL absorption using the model of \citet{finke10}.}
\label{SED_fig}
\vspace{1.2mm}
\end{figure}

The observed variability time-scale of $\sim$1 day (Fig.~\ref{LAT}) constrains the size of the emitting region, consistent with the $\gamma$-ray
light curve (Fig.~\ref{MWL}).  The results of the modelling can be found in Table
\ref{table_fit} and Figure \ref{SED_fig} \citep[for a description of the
model parameters see][]{dermer09_EC}.  Both models have the jet power in magnetic
field about a factor of 10 above the jet power in electrons\footnote{Jet
  powers were calculated assuming a two-sided jet. See
  \citet{finke08_SSC}.}. However, it is possible that the combined power in electrons and protons in the jet could be in
equipartition with the magnetic field.  If this is the case, it implies that
the jet power in protons is $P_{j,p} \approx 4.8\times10^{45}$\ erg s$^{-1}$
from the 2011 model and $P_{j,p} \approx 1.9\times10^{46}$\ erg s$^{-1}$ from
the 2013 flaring model. This model assumes the emitting region is outside the BLR, where dust torus photons are likely the seed photon source.
This source was modeled as being
an isotropic, monochromatic radiation source with dust parameters chosen to be consistent with the relation between inner radius, disc luminosity, and dust temperature from \citet{nenkova08}. Note that the model parameters shown here are not unique, and other model
parameters could also reproduce the SED, especially for the 2011 state.

In 2011 May-June the $\gamma$-ray and X-ray flux was intermediate between the 2013
flaring activity and the low activity state observed e.g. in 2009 May 15
\citep{foschini12}, with no corresponding flare in the optical. This aspect and the
shape of the optical spectrum indicate that the optical emission during that
period was dominated by thermal accretion disc emission. The synchrotron
component was poorly constrained, other than an upper limit from the optical
data. According to our model, thermal emission from the dust torus would be visible if
infrared observations were available. The X-rays originated from a combination
of coronal emission (for the soft excess) and EC emission.  In this model, the
SSC emission was quite weak and did not play a significant role in the SED.

The 2013 activity state was brighter across the electromagnetic spectrum than during 2011. In particular, the 2013 flaring period included a flare in
optical as well as in X-rays and $\gamma$ rays. The much stronger optical emission
implies that the two activity states differed by more than just the electron
distribution. This is in contrast to the FSRQ PKS 0537$-$441, where high and
low activity states were modelled by only changing the electron distribution
\citep{dammando13b}. In this way, PMN J0948$+$0022 is similar to the FSRQ PKS 2142$-$75 \citep{dutka13_2142}, which exhibited multiple
flares where another parameter besides the electron distribution needed to be
changed between flares. As for PKS 2142$-$75, we modelled the two activity
states of PMN J0948$+$0022 by changing the magnetic field and the electron distribution. This may be an indication that the shock responsible for the flare produces a compression as it passes through the emitting region yielding
an amplification of the magnetic field during the flare. The 2011 intermediate state from PMN J0948$+$0022
is similar to ``flare B'' from PKS 2142$-$75 \citep{dutka13_2142}, since in
both cases the optical part of the spectrum is dominated by the thermal disc emission.
The 2013 flare from PMN J0948$+$0022 is most like ``flare A'' from PKS
2142$-$75 \citep{dutka13_2142}, since for both of these flares the optical
spectrum of the two sources is dominated by non-thermal synchrotron
emission. We also modeled the 2013 flaring state with parameters consistent with
Compton-scattering of BLR H$\alpha$ (6563 \AA) line radiation; again,
see Figure 4 and Table 5.  The model reproduces the data approximately
equally as well as the dust scattering model.  Thus, it is not
possible to distinguish between the two scenarios by snapshot spectral
modeling.  However, we note the BLR scattering model is farther from
equipartition, so this may be a weak argument in favor of the dust
scattering scenario.
As also observed in the NLSy1 SBS 0846+513 \citep{dammando12}, PMN
J0948$+$0022 exhibits behavior similar to other FSRQs. 
 The X-ray emission is dominated by EC, and the coronal emission is not visible, so that there is no
evidence for a soft excess from the data in 2013. However, this could be due to the low sensitivity of {\em Swift}-XRT at $\sim$1 keV.

\begin{figure}
\vspace{2.2mm} 
\includegraphics[width=8cm]{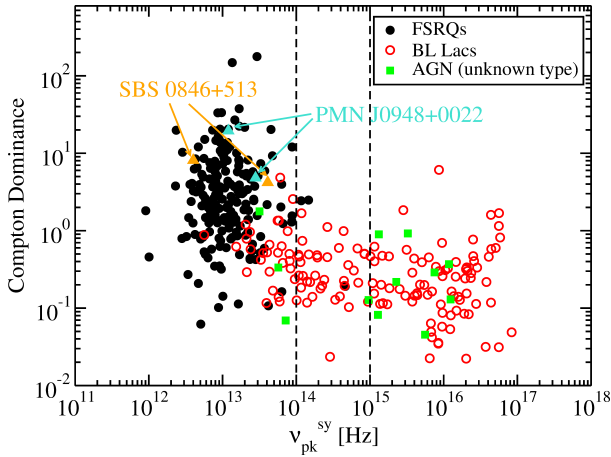}
\caption{The Compton dominance versus peak synchrotron frequency for
sources from the 2LAC \citep{ackermann11}, along with two states
of SBS 0846$+$513 \citep{dammando13a} and PMN J0948$+$0022 (this work).
Filled circles represent FSRQs, empty circles represent BL Lac
objects, and filled squares represent AGN of unknown type. Adapted
from \citet{finke13}. PMN J0948$+$0022 and SBS 0846$+$513 are plotted as triangles.}
\label{CD_fig}
\vspace{2.2mm}
\end{figure}

A clear correlation between Compton dominance\footnote{Compton dominance is the ratio of the peak Compton luminosity to peak synchrotron luminosity.} and the rest-frame peak synchrotron frequency ($\nu_{pk}^{sy}$) for blazars was reported in \citet{finke13}. This correlation is related to the
contribution of the EC component, which results to be higher for larger values of Compton
dominance. In Fig.~\ref{CD_fig}, the Compton dominance versus $\nu_{pk}^{sy}$ is
shown for sources from the second LAT AGN Catalog
\citep[2LAC;][]{ackermann11}. In addition we plot the low and high activity
states of the NLSy1 SBS 0846+513 \citep{dammando13a} and the intermediate and
high activity states of PMN J0948$+$0022 shown in Fig.~\ref{SED_fig}. Both
NLSy1s are in a region occupied by FSRQs, indicating the
similarity between FSRQs and these NLSy1s. It is worth mentioning that for both NLSy1s the peak frequency is higher for lower Compton
dominances, in agreement with the general behaviour observed for blazars.

\begin{table*}
\footnotesize
\begin{center}
\caption{Model parameters for the SEDs shown in Fig.~\ref{SED_fig}.}
\label{table_fit}
\begin{tabular}{lcccc}
\hline
Parameter & Symbol & 2011 intermediate & 2013 flare (dust torus) &  2013 flare
(BLR) \\
\hline
Redshift & 	$z$	& 0.585 & 0.585	& 0.585 \\
Bulk Lorentz Factor & $\Gamma$	& 30 & 30 & 12 \\
Doppler factor & $\delta_D$	& 30 & 30 & 12 \\
Magnetic Field (G) & $B$         & 0.5 & 1.0  & 1.0 \\
Variability Time-Scale (s) & $t_v$       & $10^5$ & $10^5$  & $10^5$ \\
Comoving radius of blob (cm) & $R^{\prime}_b$ & 5.7$\times$10$^{16}$ &
5.7$\times$10$^{16}$ & $2.3\times10^{16}$ \\
\hline
Low-Energy Electron Spectral Index & $p_1$       & 2.5 & 2.35 & 2.35 \\
High-Energy Electron Spectral Index  & $p_2$       & 4.0 & 4.0 & 3.9 \\
Minimum Electron Lorentz Factor & $\gamma^{\prime}_{min}$  & 2.7 & 1.0 & 1.0\\
Break Electron Lorentz Factor & $\gamma^{\prime}_{brk}$ & $5.0\times10^2$ & $6.8\times10^2$ & $7.8\times10^2$ \\
Maximum Electron Lorentz Factor & $\gamma^{\prime}_{max}$  & $1.0\times10^4$ & $1.0\times10^4$  & $1.0\times10^4$\\
\hline
Black hole Mass (M$_\odot$) & $M_{BH}$ & $1.5\times10^8$ & $1.5\times10^8$  & $1.5\times10^8$\\
Disc luminosity (erg s$^{-1}$) & $L_{disc}$ & $5.7\times10^{45}$ & $5.7\times10^{45}$ & $5.7\times10^{45}$\\
Inner disc radius (R$_{\rm g}$) & $R_{in}$ & $6.0$ & $6.0$ & $6.0$\\
Accretion efficiency     & $\eta_{disc}$ & 1/12 & 1/12 & 1/12\\ 
Fraction of photons scattered by corona & $f_{sc}$ & 0.35 & 0.35 & 0.35\\
Photon index of scattered coronal emission & $\Gamma_{sc}$ & 2.3 & 2.3 & 2.3\\
\hline
Seed photon source energy density (erg cm$^{-3}$) & $u_{seed}$ & $2.0\times10^{-4}$ & $2.0\times10^{-4}$ & 0.3 \\
Seed photon source photon energy (m$_e$ c$^2$) & $\epsilon_{seed}$ & $7.0\times10^{-7}$ & $7.0\times10^{-7}$ & $3.7\times10^{-6}$ \\
Dust Torus luminosity (erg s$^{-1}$) & $L_{dust}$ & $1.1\times10^{45}$ & $1.1\times10^{45}$ & $1.1\times10^{45}$ \\
Dust Torus radius (cm) & $R_{dust}$ & $3.8\times10^{18}$ & $3.8\times10^{18}$ & $3.8\times10^{18}$ \\
Dust temperature (K) & $T_{dust}$ & $1700$  & $1700$ &  $1700$ \\
\hline
Jet Power in Magnetic Field (erg s$^{-1}$) & $P_{j,B}$ & $5.4\times10^{45}$ & $2.2\times10^{46}$ & $5.5\times10^{46}$  \\
Jet Power in Electrons (erg s$^{-1}$) & $P_{j,e}$ & $6.3\times10^{44}$ & $2.4\times10^{45}$ & $1.5\times10^{44}$ \\
\hline
\end{tabular}
\end{center}
\end{table*}

\section{Conclusions}

We reported on the observation by the {\em Fermi}-LAT of flaring $\gamma$-ray
activity from the NLSy1 PMN J0948$+$0022 in 2012 December--2013 January. The
$\gamma$-ray flux reached on 2013 January 1, (155 $\pm$ 31) $\times$10$^{-8}$ ph
cm$^{-2}$ s$^{-1}$ in the 0.1--100 GeV energy range, is the highest value
detected from this source and more generally from a NLSy1 so far. Only marginal spectral variability was
observed in $\gamma$ rays during the flare. PMN J0948$+$0022 was not detected
above E $>$ 10 GeV by LAT.

\noindent We presented multiwavelength observations of the source from radio
to VHE during the period 2012 December 1--2013 August 31 including data
collected by VERITAS, {\em Fermi}-LAT, {\em Swift}, CRTS, MOJAVE, and OVRO. An
intense flaring episode, quasi-simultaneous to the $\gamma$-ray one, was
observed on 2012 December 30 in optical, UV, and X-rays, showing record flux also in these energy bands. This suggests a strict connection between the
region responsible for the $\gamma$-ray activity and that responsible for the
optical-to-X-ray activity, although the time sampling of the observations is
too sparse from optical to X-rays to formally establish a correlation. Flaring
activity was also observed at 15 GHz with the radio peak detected on
2013 June 2. The most likely scenario is that the radio flare is the delayed counterpart of the $\gamma$-ray one, with a delay of $\sim$5
months due to opacity effects and the propagation of the shock along the jet. Past flaring episodes from PMN J0948$+$0022
have shown a different delay or no indication of correlated variability \citep{foschini11,dammando14}, thus
making the radio-gamma connection complex in this object.
A slightly higher polarized emission from the core and polarization percentage was observed by MOJAVE after the peak of the radio emission, while no significant change of the EVPA was detected.

Following the $\gamma$-ray flare observed by the LAT, a first detection of VHE emission from PMN J0948$+$0022 was attempted by VERITAS. The observations carried out between 2013 January 6 and 17, just at
the end of the $\gamma$-ray flaring activity, resulted in an upper limit of
$F_{>0.2\,\mathrm{TeV}} < 4.0 \times10^{-12}\,\mathrm{ph}\,\mathrm{cm}^{-2}\,\mathrm{s}^{-1}$. Further observations with the current
generation of IACTs (VERITAS, MAGIC, and H.E.S.S.), especially during the
peak of $\gamma$-ray activity, may lead to VHE detections of this object. With a lower energy
threshold and better sensitivity, future observations with the Cherenkov
Telescope Array will constrain the level of $\gamma$-ray emission at 100 GeV
and below \citep{sol13}. However, pair production due to the EBL significantly
suppresses the VHE flux of PMN J0948$+$0022 given its redshift of 0.5846. 1H
0323$+$342 is another, closer ($z$ = 0.061), NLSy1, already observed (but not significantly detected) at VHE
during a search for TeV-emitting FSRQs \citep{falcone04}. If 1H 0323$+$342 exhibits flaring behavior similar to that observe
d in PMN J0948+0022 then it will be a promising target for IACTs.

We compared the broad-band SED of the 2013 flaring activity state with that from an
intermediate activity state of PMN J0948$+$0022 observed in 2011. Contrary to
what was observed for some FSRQs \citep[e.g., PKS 0537$-$441;][]{dammando13b} the two SEDs, modelled as
an EC component of seed photons from a dust torus, could not be modeled by
changing only the electron distribution parameters. A higher magnetic field is
needed for modelling the high activity state of PMN J0948$+$0022, as was also
observed for the FSRQ PKS 2142$-$75 \citep{dutka13_2142}. In a Compton dominance vs. synchrotron peak frequency plot (see
Figure~\ref{CD_fig}), the values for the two activity states of PMN J0948$+$0022 lie
in the same region occupied by the FSRQs, as well as the other NLSy1 SBS
0846$+$513. This suggests that the EC scattering is likely the dominant mechanism for
producing high-energy emission in PMN J0948$+$0022, confirming the
similarities between $\gamma$-ray emitting NLSy1s and FSRQs. The inverse 
Compton (IC) peak of this source seems to lie at 100 MeV or below. With the improved
sensitivity of the LAT at low energies with Pass 8 data \citep{atwood13} we
will be able to characterize in more detail the IC bump of PMN J0948$+$0022,
and therefore the SED modelling. 

In addition to the large uncertainties on the BH mass of the NLSy1
\citep[e.g.,][]{marconi08, calderone13}, the structure of the host galaxies is
unclear. For four out of five NLSy1s detected
in $\gamma$ rays by LAT, no high-resolution observations of their host galaxies
are available. This is also related to the redshifts of these objects, and then
the difficulty of detecting significant resolved structures for their host
galaxies \citep[e.g., SBS 0846$+$513;][]{maoz93}. Therefore we cannot rule out the possibility that these $\gamma$-ray emitting NLSy1s are hosted in elliptical galaxies.

\section*{Acknowledgements}

The {\em Fermi} LAT Collaboration acknowledges generous ongoing
support from a number of agencies and institutes that have supported
both the development and the operation of the LAT as well as
scientific data analysis.  These include the National Aeronautics and
Space Administration and the Department of Energy in the United
States, the Commissariat \`a l'Energie Atomique and the Centre
National de la Recherche Scientifique / Institut National de Physique
Nucl\'eaire et de Physique des Particules in France, the Agenzia
Spaziale Italiana and the Istituto Nazionale di Fisica Nucleare in
Italy, the Ministry of Education, Culture, Sports, Science and
Technology (MEXT), High Energy Accelerator Research Organization (KEK)
and Japan Aerospace Exploration Agency (JAXA) in Japan, and the
K.~A.~Wallenberg Foundation, the Swedish Research Council and the 
Swedish National Space Board in Sweden. Additional support for science
analysis during the operations phase is gratefully acknowledged from
the Istituto Nazionale di Astrofisica in Italy and the Centre National
d'\'Etudes Spatiales in France.

The VERITAS Collaboration is grateful to Trevor Weekes for his seminal
contributions and leadership in the field of VHE gamma-ray astrophysics,
which made this study possible.
The work of the VERITAS Collaboration is supported by grants from the U.S. Department of Energy Office
of Science, the U.S. National Science Foundation and the Smithsonian Institution, by NSERC in Canada, by
Science Foundation Ireland (SFI 10/RFP/AST2748) and by the Science and Technology Facilities
Council in the U.K. We acknowledge the excellent work of the technical support
staff at the Fred Lawrence Whipple Observatory and at the collaborating institutions in the construction and operation of the instrument.

We thank the {\em Swift} team for making these observations possible, the
duty scientists, and science planners. The OVRO 40-m monitoring program
is supported in part by NASA grants NNX08AW31G and NNX11A043G, and NSF grants AST-0808050 
and AST-1109911. The CRTS survey is supported by the U.S.~National Science Foundation under
grants AST-0909182. This research has made use of data from the MOJAVE
database that is maintained by the MOJAVE team (Lister et al. 2009, AJ, 137,
3718). Data from the Steward Observatory spectropolarimetric monitoring
project were used. This program is supported by Fermi Guest Investigator
grants NNX08AW56G, NNX09AU10G, and NNX12AO93G. The National Radio Astronomy Observatory is a facility of the National Science
Foundation operated under cooperative agreement by Associated Universities,
Inc. We thank F. Schinzel, S. Digel, P. Bruel, and the referee, Anthony M. Brown, for useful comments and suggestions.

\vspace*{0.5cm}

\noindent
$^{1}$Dipartimento di Fisica e Astronomia, Universit\'a degli Studi di
Bologna, Viale B. Pichat, 6/2, I-40127, Bologna, Italy \\
$^{2}$INAF - Istituto di Radioastronomia, Via Gobetti 101, I-40129 Bologna, Italy\\
$^{3}$U.S. Naval Research Laboratory, Code 7653, 4555 Overlook Ave. SW, Washington, DC 20375-5352, USA \\
$^{4}$INAF - Osservatorio Astrofisico di Torino, Via Osservatorio 20, I-10025 Pino Torinese (TO), Italy \\
$^{5}$Cahill Center for Astronomy and Astrophysics, California Institute of Technology 1200 E. California Blvd., Pasadena, CA 91125, USA \\
$^{6}$KTH, Department of Physics, and the Oskar Klein Centre, AlbaNova, SE-106 91 Stockholm, Sweden \\
$^{7}$National Radio Astronomy Observatory (NRAO), P.O. Box 0, Socorro, NM 87801, USA \\
$^{8}$NASA Goddard Space Flight Center, Greenbelt, MD 20771, USA \\
$^{9}$Department of Physics, Purdue University, Northwestern Avenue 525, West Lafayette, IN 47907, USA \\
$^{10}$Department of Physics, Washington University, St. Louis, MO 63130, USA \\
$^{11}$Fred Lawrence Whipple Observatory, Harvard-Smithsonian Center for Astrophysics, Amado, AZ 85645, USA \\
$^{12}$Department of Physics and Astronomy and the Bartol Research Institute, University of Delaware, Newark, DE 19716, USA \\
$^{13}$School of Physics, University College Dublin, Belfield, Dublin 4, Ireland \\
$^{14}$Department of Physics and Astronomy, Iowa State University, Ames, IA 50011, USA \\
$^{15}$Institute of Physics and Astronomy, University of Potsdam, 14476 Potsdam-Golm, Germany \\
$^{16}$DESY, Platanenallee 6, 15738 Zeuthen, Germany \\
$^{17}$Astronomy Department, Adler Planetarium and Astronomy Museum, Chicago, IL 60605, USA \\
$^{18}$Department of Physics and Astronomy, Barnard College, Columbia University, NY 10027, USA \\
$^{19}$Department of Astronomy and Astrophysics, 525 Davey Lab, Pennsylvania State University, University Park, PA 16802, USA \\ 
$^{20}$School of Physics and Astronomy, University of Minnesota, Minneapolis, MN 55455, USA \\
$^{21}$Santa Cruz Institute for Particle Physics and Department of Physics, University of California, Santa Cruz, CA 95064, USA \\
$^{22}$School of Physics, National University of Ireland Galway, University Road, Galway, Ireland \\ 
$^{23}$Department of Physics and Astronomy, University of Iowa, Van Allen Hall, Iowa City, IA 52242, USA \\
$^{24}$Physics Department, Columbia University, New York, NY 10027, USA \\
$^{25}$Department of Physics and Astronomy, University of Utah, Salt Lake City, UT 84112, USA \\
$^{26}$Department of Physics and Astronomy, DePauw University, Greencastle, IN 46135-0037, USA \\
$^{27}$Kavli Institute for Cosmological Physics, University of Chicago, Chicago, IL 60637, USA \\
$^{28}$School of Physics and Center for Relativistic Astrophysics, Georgia Institute of Technology, 837 State Street NW, Atlanta, GA 30332-0430 \\
$^{29}$Department of Physics and Astronomy, University of California, Los Angeles, CA 90095, USA \\
$^{30}$Physics Department, McGill University, Montreal, QC H3A 2T8, Canada \\
$^{31}$Department of Applied Physics and Instrumentation, Cork Institute of Technology, Bishopstown, Cork, Ireland \\
$^{32}$Enrico Fermi Institute, University of Chicago, Chicago, IL 60637, USA \\
$^{33}$Argonne National Laboratory, 9700 S. Cass Avenue, Argonne, IL 60439, USA
\label{lastpage}


\begin{thebibliography}{99}

\bibitem[Abdo et al.(2009a)]{abdo09a} Abdo, A. A., et al. 2009a, ApJ, 699, 976

\bibitem[Abdo et al.(2009b)]{abdo09b} Abdo, A. A., et al. 2009b, ApJ, 707, 727

\bibitem[Abdo et al.(2009c)]{abdo09c} Abdo, A. A., et al. 2009c, ApJ, 707, L142

\bibitem[Abramowski et al.(2013)]{abramowski13} Abramowski, A., et al. 2013, A\&A, 554, 107

\bibitem[Acciari et al.(2008)]{ver_analysis2} Acciari, V.~A., et al.\ 2008, ApJ, 679, 1427 

\bibitem[Ackermann et al.(2011)]{ackermann11} Ackermann, M., et al.\ 2011, ApJ, 743, 171

\bibitem[Ackermann et al.(2012)]{ackermann12} Ackermann, M., et al. 2012, ApJ, 747, 104

\bibitem[Ackermann et al.(2012)]{ackermann12b} Ackermann, M., et al. 2012b, ApJS, 203, 4 

\bibitem[Albert et al.(2008)]{albert08} Albert, J., et al. 2008, Science, 320, 1752

\bibitem[Aleksic et al.(2011)]{aleksic11} Aleksic, J., et al. 2011, ApJ, 730, L8

\bibitem[Aleksi{\'c} et al.(2014)]{aleksic14} Aleksi{\'c}, J., et al.\ 2014, A\&A, 569, 46

\bibitem[Archambault et al.(2013)]{ver_analysis} Archambault, S., et al.\ 2013, ApJ, 776, 69 

\bibitem[Atwood et al.(2009)]{atwood09} Atwood, W. B., et al. 2009, ApJ, 697, 1071

\bibitem[Atwood et al.(2013)]{atwood13} Atwood, W. B., et al. 2013, 2012 Fermi Symposium proceedings - eConf C121028 (arXiv:1303.3514)

\bibitem[Baars et al.(1977)]{baars77} Baars, J. W. M., Genzel, R., Pauliny-Toth, I. I. K, Witzel, A. 1977, A$\&$A, 61, 99 

\bibitem[Barthelmy et al.(2005)]{barthelmy05} Barthelmy, S. D., et al. 2005, SSRv, 120, 143 

\bibitem[Baumgartner et al.(2013)]{baumgartner13} Baumgartner, W. H., et al. 2013, ApJS, 207, 19

\bibitem[Berge et al.(2007)]{refl} Berge, D., Funk, S., \& Hinton, J.\ 2007, A\&A, 466, 1219 

\bibitem[Bertin \& Arnouts(1996)]{bertin96} Bertin, E., \& Arnouts, S. 1996, A\&AS,  117, 393

\bibitem[Brown(2013)]{brown13} Brown, A.~M.\ 2013, MNRAS, 431, 824

\bibitem[Burrows et al.(2005)]{burrows05} Burrows, D. N., et al. 2005, SSRv, 120,165  

\bibitem[Calderone et al.(2013)]{calderone13} Calderone, G., Ghisellini, G., Colpi, M., Dotti, M. 2013, MNRAS, 431, 210

\bibitem[Cardelli et al.(1989)]{cardelli89} Cardelli, J. A., Clayton, G. C., Mathis, J. S. 1989, ApJ, 345, 245 

\bibitem[Carpenter et al.(2013)]{carpenter13} Carpenter, B., \& Ohja, R. 2013, The Astronomer's Telegram, 5344

\bibitem[Cash(1979)]{cash79} Cash, W. 1979, ApJ, 228, 939

\bibitem[D'Ammando \& Ciprini(2011)]{dammando11} D'Ammando, F., \& Ciprini, S.\ 2011, The Astronomer's Telegram, 3429 

\bibitem[D'Ammando et al.(2012)]{dammando12} D'Ammando, F., et al. 2012, MNRAS, 426, 317 

\bibitem[D'Ammando et al.(2013a)]{dammando13a} D'Ammando, F., et al. 2013a, MNRAS, 436, 191

\bibitem[D'Ammando et al.(2013b)]{dammando13b} D'Ammando, F., et al.\ 2013b, MNRAS, 431, 2481 

\bibitem[D'Ammando \& Orienti(2013c)]{dammando13c} D'Ammando, F., \& Orienti, M.\ 2013c, The Astronomer's Telegram, 4694 

\bibitem[D'Ammando et al.(2014)]{dammando14} D'Ammando, F., et al. 2014, MNRAS, 438, 3521

\bibitem[Dermer et al.(2009)]{dermer09_EC} Dermer, C.~D., Finke, J.~D., Krug, H., B\"ottcher, M.\ 2009, ApJ, 692, 32

\bibitem[Dermer et al.(2014)]{dermer14} Dermer, C.~D., Cerruti, M., Lott, B., Boisson, C., Zech, A.\ 2014, ApJ, 782, 82

\bibitem[Djorgovski et al.(2011)]{djorgovski11} Djorgovski, S.~G., et al. 2011, in The First Year of MAXI: Monitoring Variable X-ray Sources, eds. T. Mihara \& N. Kawai, Tokyo: JAXA Special Publ. [arXiv:1102.5004]

\bibitem[Dominguez et al.(2011)]{dominguez11} Dominguez, A., et al. 2011, MNRAS, 410, 2556

\bibitem[Donato(2010)]{donato10} Donato, D. 2010, The Astronomer's Telegram, 2733

\bibitem[Donato(2011)]{donato11} Donato, D. 2011, The Astronomer's Telegram, 3452

\bibitem[Drake et al.(2009)]{drake09} Drake, A. J., et al. 2009, ApJ, 696, 870

\bibitem[Dutka et al.(2013)]{dutka13_2142} Dutka, M.~S., et al. 2013, ApJ, 779, 174 

\bibitem[Eggen et al.(2013)]{eggen13} Eggen, J. R., Miller, H. R., Maune, J. D. 2013, ApJ, 773, 85

\bibitem[Falcone et al.(2004)]{falcone04} Falcone, A.~D., et al. 2004, ApJ, 613

\bibitem[Finke et al.(2008)]{finke08_SSC} Finke, J.~D., Dermer, C.~D., B\"ottcher, M.\ 2008, ApJ, 686, 181

\bibitem[Finke et al.(2010)]{finke10} Finke, J.~D., Razzaque, S., Dermer, C.~D. 2010, ApJ, 712, 238

\bibitem[Finke(2013a)]{finke13} Finke, J.~D.\ 2013a, ApJ, 763, 134

\bibitem[Finke(2013b)]{finke13_review} Finke, J.\ 2013b, arXiv:1303.5095

\bibitem[Fomin et al.(1994)]{fomin} Fomin, V.~P., et al.\ 1994, Astroparticle Physics, 2, 137
 
\bibitem[Foschini(2010)]{foschini10} Foschini, L. 2010, The Astronomer's Telegram, 2752

\bibitem[Foschini et al.(2011)]{foschini11} Foschini, L., et al.\ 2011, MNRAS, 413, 1671

\bibitem[Foschini et al.(2012)]{foschini12} Foschini, L., et al. 2012, A$\&$A, 548, 106

\bibitem[Gehrels et al.(2004)]{gehrels04} Gehrels, N., et al. 2004, ApJ, 611, 1005 

\bibitem[Giroletti et al.(2011)]{giroletti11} Giroletti, M., et al. 2008, A\&A, 528, L11

\bibitem[Hillas et al.(1998)]{hillas-crab} Hillas, A.~M., et al.\ 1998, ApJ, 503, 744

\bibitem[Holder et al.(2008)]{VERITAS} Holder, J., et al.\ 2008, American Institute of Physics Conference Series, 1085, 657

\bibitem[Holder et al.(2006)]{VERITAST1} Holder, J., et al.\ 2006, Astroparticle Physics, 25, 391

\bibitem[Kalberla et al.(2005)]{kalberla05} Kalberla, P. M. W., Burton, W. B., Hartmann, D., Arnal, E. M., Bajaja, E., Morras, R., P\"{o}ppel, W. G. L. 2005, A$\&$A, 440, 775

\bibitem[Kieda et al.(2013)]{kieda} Kieda, D.~B.~ for the VERITAS Collaboration\ 2013, in Proc. 33rd Int. Cosmic Ray Conf. (Rio de Janeiro), arXiv:1308.4849

\bibitem[K{\"o}nigl(1981)]{konigl81} Konigl, A.\ 1981, ApJ, 243, 700

\bibitem[Li \& Ma(1983)]{lima} Li, T.-P., \& Ma, Y.-Q.\ 1983, ApJ, 272, 317

\bibitem[Lister et al.(2009)]{lister09} Lister, M. L., et al. 2009, AJ, 137, 3718

\bibitem[Lister et al.(2013)]{lister13} Lister, M. L., et al. 2013, AJ, 146, 120

\bibitem[Maoz et al.(1993)]{maoz93} Maoz, D., Bahcall, J. N., Doxsey, R., Schneider, D. P., Bahcall, N. A., Lahav, O., Yanny, B. 1993, ApJ, 402, 69
 
\bibitem[Marconi et al.(2008)]{marconi08} Marconi, A., Axon, D., Maiolino, R., Nagao, T., Pastorini, G., Pietrini, P., Robinson, A., Torricelli, G. 2008, ApJ, 678, 693

\bibitem[Marscher et al.(2010)]{marscher10} Marscher, A.~P., et al.\ 2010, ApJL, 710, 126

\bibitem[Marscher(2014)]{marscher14} Marscher, A. P. 2014, ApJ, 780, 87

\bibitem[Mattox et al.(1996)]{mattox96} Mattox, J. R., et al. 1996, ApJ, 461, 396 

\bibitem[Nalewajko et al.(2014)]{nalewajko14} Nalewajko, K., Begelman, M. C., Sikora, M.\ 2014, ApJ, 789, 161

\bibitem[Nenkova et al.(2008)]{nenkova08} Nenkova, M., Sirocky, M.~M., Nikutta, R., Ivezi{\'c}, {\v Z}., \& Elitzur, M.\ 2008, ApJ, 685, 160 

\bibitem[Nolan et al.(2012)]{nolan12} Nolan, P., et al. 2012, ApJS, 199, 31 

\bibitem[Orienti et al.(2013)]{orienti13} Orienti, M., et al. 2013, MNRAS, 428, 2418

\bibitem[Osterbrock $\&$ Pogge(1985)]{osterbrock85} Osterbrock, D. E., $\&$ Pogge, R. W. 1985, ApJ, 297, 166 

\bibitem[Pacciani et al.(2014)]{pacciani14} Pacciani, L., et al.\ 2014, ApJ, 790, 45

\bibitem[Perlman(2000)]{perlman00} Perlman, E. S. 2000, AIPC, 515, 53

\bibitem[Pogge(2000)]{pogge00} Pogge, R. W. , New Astronomy Reviews 2000, 44, 381 

\bibitem[Poutanen \& Stern(2010)]{poutanen10} Poutanen, J., \& Stern, B.\ 2010, ApJL, 717, 118

\bibitem[Raiteri et al.(2011)]{raiteri11} Raiteri, C. M., et al. 2011, A$\&$A, 534, 87 

\bibitem[Raiteri et al.(2012)]{raiteri12} Raiteri, C. M., et al. 2012, A\&A, 545, 48

\bibitem[Richards et al.(2011)]{richards11} Richards, J. L., et al. 2011, ApJS, 194, 29 

\bibitem[Rolke et al.(2005)]{rolke} Rolke, W.~A., L{\'o}pez, A.~M., \& Conrad, J.\ 2005, Nuclear Instruments and Methods in Physics Research A, 551, 493 

\bibitem[Roming et al.(2005)]{roming05} Roming, P. W. A., et al. 2005, SSRv, 120, 95 

\bibitem[Saito et al.(2013)]{saito13} Saito, S., et al.\ 2013, ApJL, 766, 11

\bibitem[Schlafly \& Finkbeiner(2011)]{schlafly11} Schlafly, E. F. \& Finkbeiner, D. P. 2011, ApJ, 737, 103

\bibitem[Schneider et al.(2010)]{schneider10} Schneider, D. P., et al. 2010, AJ, 139, 2360

\bibitem[Sol et al.(2013)]{sol13} Sol, H., et al. 2013, APh, 43, 215

\bibitem[Steiner et al.(2009)]{steiner09} Steiner, J.~F., Narayan, R., McClintock, J.~E., \& Ebisawa, K.\ 2009, PASP, 121, 1279

\bibitem[Stern \& Poutanen(2011)]{stern11} Stern, B.~E., \& Poutanen, J.\ 2011, MNRAS, 417, L11

\bibitem[Stern \& Poutanen(2014)]{stern14} Stern, B.~E., \& Poutanen, J.\ 2014, arXiv:1408.0793

\bibitem[Tavecchio et al.(2010)]{tavecchio10} Tavecchio, F., et al. \ 2010, MNRAS, 405, L94

\bibitem[Urry \& Padovani(1995)]{urry95} Urry, M., \& Padovani, P. 1995, PASP, 107, 803

\bibitem[Wilms et al.(2000)]{wilms00} Wilms, J., Allen, A., McCray, R. 2000, ApJ, 542, 914 

\bibitem[Zhou et al.(2003)]{zhou03} Zhou, H.-Y., Wang, T.-G., Dong, X.-B., Zhou, Y.-Y., \& Li, C.\ 2003, ApJ, 584, 147 

\end{thebibliography}
\end{document}